\documentclass[11pt,twoside]{article}

\usepackage{asp2004}
\usepackage{epsf}
\usepackage{psfig}
\usepackage{lscape}

\begin{document}
\title{Solid state components of varying composition in the outflow of the Red Rectangle}   
\author{F.~Kemper}   
\affil{University of Virginia, PO Box 3818, Department of Astronomy, Charlottesville, VA 22903-0818}    
\author{J.D.~Green}
\affil{ Department of Physics and Astronomy, University of Rochester, Rochester, NY 14627}
\author{E.~Peeters}
\affil{NASA Ames Research Center, MS 245-6, Moffett Field, CA 94035}

\begin{abstract} 
We present high resolution Spitzer spectroscopy of several pointings in the
outflow of the Red Rectangle. Besides the resonances due to the C-rich PAHs, the spectra
show a wealth of new emission bands, in particular in the 13--19.5
$\mu$m range.  We argue that these bands are due to solid state
components, and show that simple oxides of a varying composition show
a good match to the data. The presence of these O-rich species in the
supposedly C-rich outflow is unexplained.
\end{abstract}

\section{Introduction}
The Red Rectangle (HD 44179) is a nearby \citep[710 pc;][]{MST_02_RR}
post-Asymptotic Giant Branch
(AGB) star. The post-AGB phase is a relatively short-lived phase
\citep[10$^4$--10$^5$ years;][]{V_03_postAGB} in the post-Main-Sequence evolution of stars
with $M<8M_{\odot}$ during which the central star sheds its dusty
circumstellar shell. The Red Rectangle is known to contain a central
binary system \citep{CAC_75_RR}. Its close distance and rare evolutionary status
makes the Red Rectangle a host for many phenomena not observed
elsewhere in the Galaxy. It is for instance the only stellar source
exhibiting the Extended Red Emission \citep[ERE;][]{VCG_02_ERE}. The
Red Rectangle is also known to show a complex circumstellar
chemistry. Using the Infrared Space Observatory (ISO)
\citet{WWV_98_RedRectangle} showed that both silicates and polycyclic
aromatic hydrocarbons (PAHs) are present in the nebula. The
oxygen-rich silicates are thought to be located in the circumbinary
disk, while the carbon-rich PAHs are found in the significantly less
dense bipolar outflows. The spectrometers onboard ISO lacked the
resolution to confirm this theory, but ground-based observations
\citep{MKO_04_RR} support the idea that the PAHs are predominantly
present in the outflow.  Imaging obtained with the Hubble Space
Telescope (HST) shows that the outflows have an inhomogeneous
structure, probably reflecting variations in the density
\citep{CVB_04_HST}. These different densities may give rise to 
a variation in dust condensates in the outflows.
We observed the northern outflow 
with the Infrared Spectrograph (IRS) onboard
the Spitzer Space Telescope to look for compositional variations in
the dust. Although
ground-based observations do have better spatial resolution, they
lack the sensitivity of the IRS. 

\section{Observations}

\begin{figure}[!ht]
\plottwo{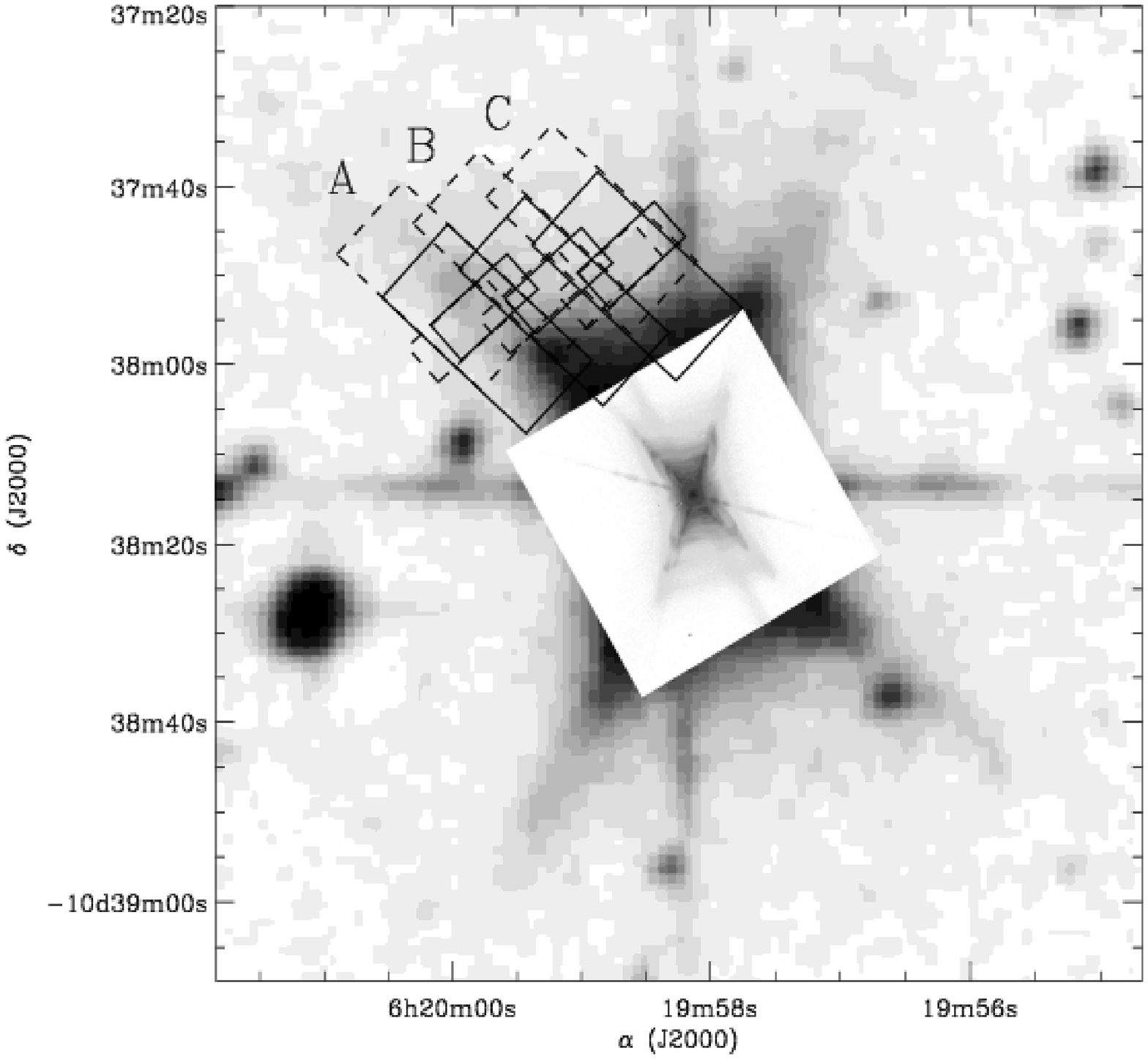}{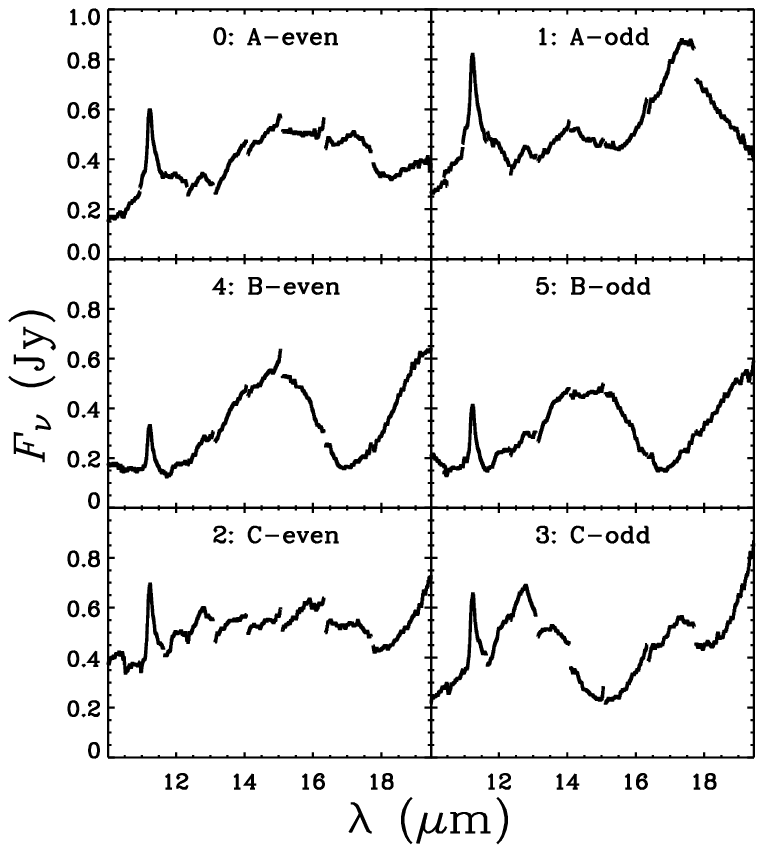}
\caption{The panel on the left hand side shows the slit positions of
the IRS high resolution modules on the outflow of the Red
Rectangle. The small boxes outline the positions of the SH module, and
the large boxes the slit of the LH module. The solid and dashed lines
indicate the two nod positions in each pointing. The positions A, B,
and C are explained in the main text. On the right hand side, full
slit spectral extractions in SH of each of the nods in A, B and C are
presented.}
\label{fig:pos}
\end{figure}

We performed IRS high resolution spectroscopy on three positions in
the outflow of the Red Rectangle. The positions were chosen in the
northern outflow, about 30$''$ away from the central star (see
Fig.~\ref{fig:pos}). Observations were done in IRS staring mode, with
integration times of 62.92 seconds for the short high (SH) observations
and 58.72 seconds for the long high (LH) observations. In this work we
will only discuss the SH data, since substantial calibration
difficulties with the LH data still exist.

The S10.5 standard pipeline data were reduced using {\sc smart} \citep{HDH_04_smart}, with the remark that extended source calibration is
still primitive.  

\section{Spectral features}

Fig.~\ref{fig:pos}
shows the full slit spectral extractions of each of the six positions,
for the SH module. The spectra are incredibly rich, with significant
variations between the different pointings. Shortwards of 13 $\mu$m
the spectra are dominated by out-of-plane (OOP) bending modes of the
PAHs \citep{HVP_01_OOPS}. Longwards
of 13 $\mu$m broad emission bands dominate the spectrum,
hereafter referred to as the MIR bands. They are
comparable in strength to the OOPs, or in some cases even stronger. It
was a surprise to see these bands, as there is no sign of
similar spectral structure in the ISO spectroscopy
\citep{WWV_98_RedRectangle}, nor do we see such bands anywhere
else. The carrier is unknown.

The MIR bands have some peculiar properties. First of all, the bands
are very strong compared to the OOP bending modes in PAHs, thus making
an identification with PAH-like species virtually
impossible. Longwards of 13 $\mu$m PAHs do show spectral features like
the PAH plateau \citep{VHP_00_PAHplateau} but these resonances are
weak compared to the OOPs.  Additional properties suggest that the
carrier of the bands is a solid state condensate: the resonances are
very wide, and the peak positions are shifting from position to
position, perhaps due to a change in grain properties or
composition. The bands are very strong compared to the continuum.

\begin{figure}[!ht]
\plottwo{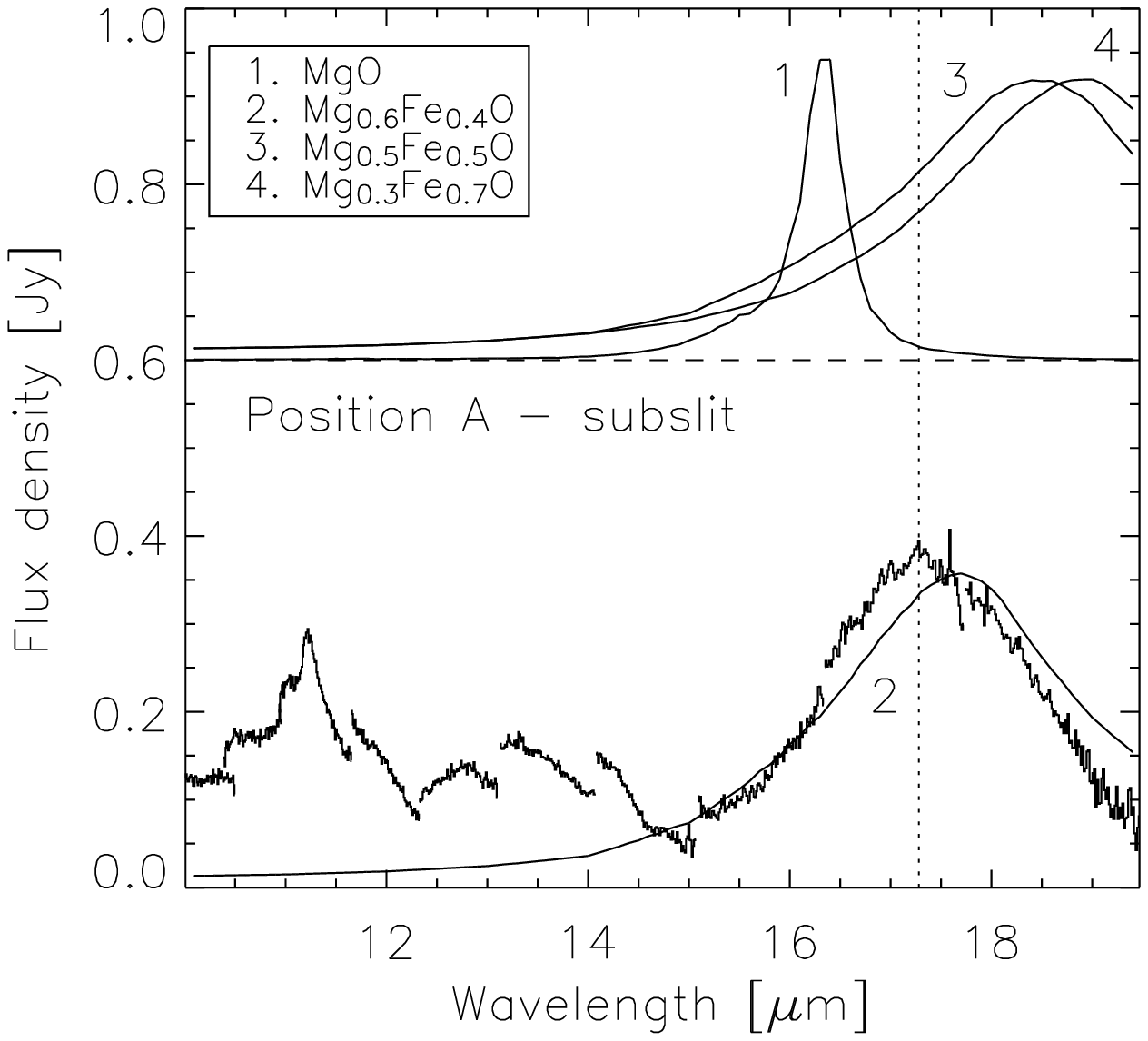}{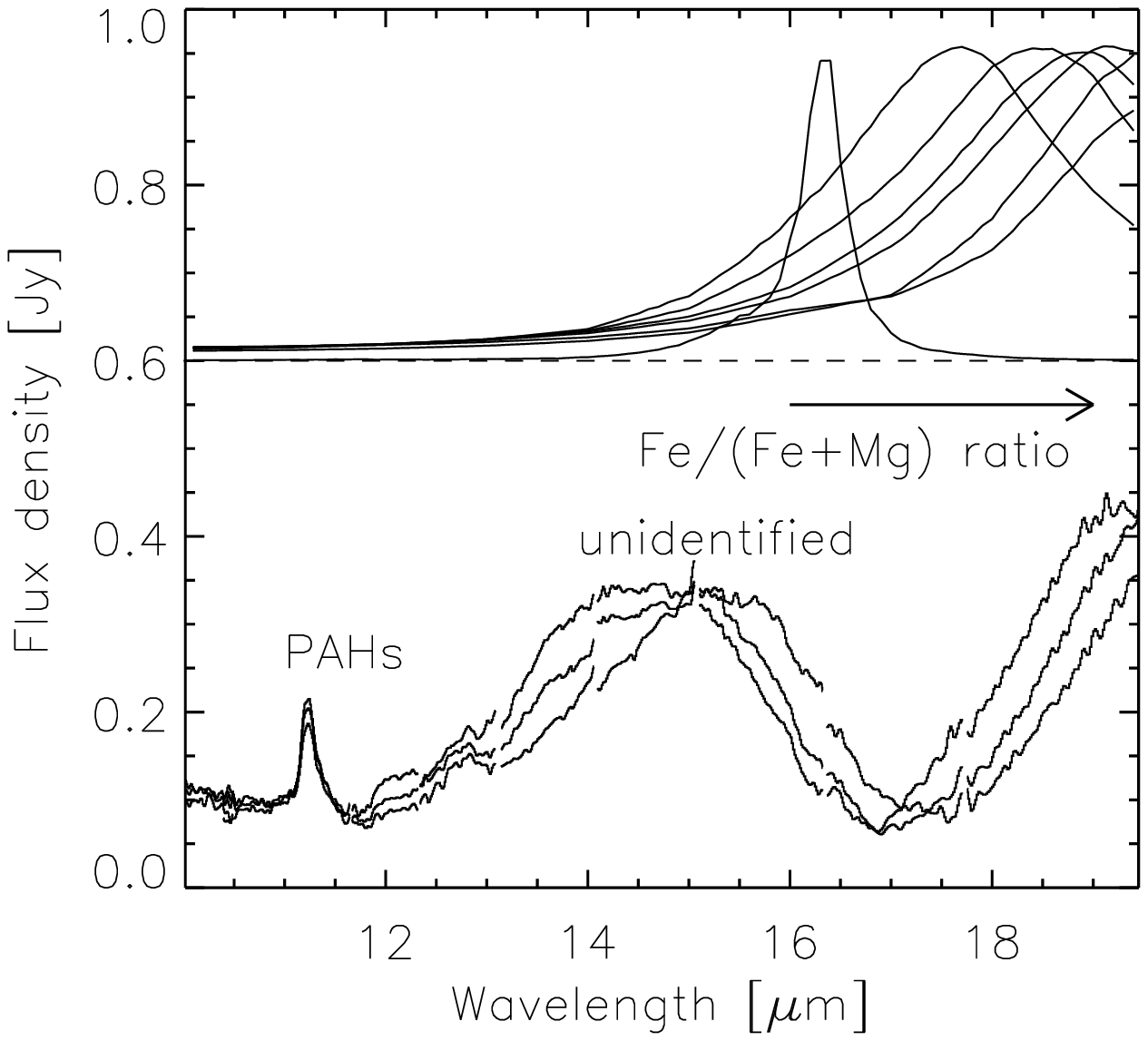}
\caption{Identification of the MIR bands. The left panel shows 
a subslit extraction of one of the nods in position A. Prominently
present is the 17.5 $\mu$m feature. Overplotted, and in the top part
of the plot are the opacities of some Fe-Mg-oxides for comparison. The
right hand panel shows subslit extractions of one of the nods of
position B. The three different curves in the lower part of the plot
indicate three equidistant subslit extractions, and the top part shows
how the 16.5 micron MgO feature shifts to longer wavelengths when iron
replaces the magnesium.}
\label{fig:ident}
\end{figure}

Not many dust components are known to show resonances in this
wavelength range. We found that besides PAHs no materials normally
found in carbon-rich environments show resonances with the properties
described above. On the other hand, some oxygen-rich components are
known to have resonances somewhere in the 13-20 $\mu$m range, which
are detected in some astrophysical environments.  A 19.5 $\mu$m
feature seen around AGB stars has been assigned to
magnesium-iron-oxides \citep{PKM_02_19.5um}, and in red giants
features at 13 and 17 $\mu$m are possibly to be due to spinels
\citep{FPM_01_spinels}.    Trend
analysis in a large number of ISO spectra shows that the observed
features are always found at the same wavelengths and are relatively
narrow \citep{SKG_03_13micron}, they do not show the large variety of
peak positions observed in the outflow of the Red Rectangle.  It turns
out that (Mg,Fe)O provides a good identification for the features,
especially longwards of 17 $\mu$m.  A variation in Fe-content explains
the shift in peak position. In the left panel of Fig.~\ref{fig:ident},
it is clear from interpolation that the strong 17.5 $\mu$m feature
observed in one of the nods of Position A can be explained with a
material slightly more Mg-rich than Mg$_{0.6}$Fe$_{0.4}$O.  In the right
hand panel, sub slit extractions in position B show that 
the Fe-content of the (Mg,Fe)O causes the resonance to shift
in peak position. The broad feature between 13--16 $\mu$m remains
unidentified, but seems to correlate with the longer wavelength (18--20 $\mu$m) 
feature in peak position shift (Fig.~\ref{fig:ident}), and could therefore perhaps  be explained by  oxide-rich composite grains. Other components
could include spinel or silicates, which could cause a shift to shorter wavelengths for the
pure oxide feature.

\section{Discussion and conclusion}

The rich variety of features longwards of 13 $\mu$m seen in the outflow of the Red Rectangle is possibly carried by
simple oxides, and composite grains containing those oxides. The
presence of such oxygen-rich species in the carbon-rich outflow  is rather surprising. Explanations may include 
that these
oxides are relicts from an earlier mass loss phase, when the outflow
was still oxygen-rich.  Alternatively the presence of these oxides may
be caused by erosion from the oxygen-rich circumbinary disk. 
The variations in the composition of the
oxides with position in the outflow is rather unusual, 
given that virtually all sources showing resonances which
have previously been identified with magnesium-iron-oxides or spinel,
do not show any variation in peak position from source to source, or
within sources \citep{SKG_03_13micron}. It is not trivial to connect
these resonances to the ones currently observed in the Red Rectangle.
Further observations of the outflow in the low resolution mode of IRS are required to gain
further insights in the distribution and the carriers of the broad MIR
bands.



\acknowledgements 
We wish to thank Mike Jura, Lou Allamandola and Xander Tielens for
inspiring discussions. Support for this work was provided by NASA
through the Spitzer Fellowship Program, under award 011 808-001.


\end{document}